\documentclass[aps,twocolumn]{revtex4}
\usepackage [dvips]{graphicx}
\def\beq{\begin{equation}}
\def\eeq{\end{equation}}
\begin{document}
\title{\bf
 SUPERFLUID PROPERTIES OF THE INNER CRUST OF NEUTRON STARS }

\author{\rm
N. Sandulescu $^{a,b,c}$
Nguyen Van Giai $^{b}$
R. J. Liotta $^{c}$ }

\bigskip

\address {\rm
  $^{a)}$~ Institute of Physics and Nuclear Engineering,
           76900 Bucharest, Romania\\
  $^{b}$~  Institut de Physique Nucl\'eaire, Universit\'e Paris-Sud,
           F-91406 Orsay Cedex, France \\
  $^{c)}$~  Royal Institute of Technology, Alba Nova,
            SE-10691, Stockholm, Sweden }

\begin{abstract}

 Superfluid properties of the inner crust matter of
 neutron stars, formed by nuclear clusters immersed in a
 dilute neutron gas, are analysed in a self-consistent HFB approach.
 The calculations are performed with two pairing forces, fixed so as
 to obtain in infinite nuclear matter the pairing gaps provided
 by the Gogny force or by induced interactions.
 It is shown that the nuclear clusters can either suppress or
 enhance the pairing correlations inside the inner crust matter,
 depending on the density of the surrounding neutrons. The profile of the
 pairing field in the inner crust is rather similar for both
 pairing forces, but the values of the pairing gaps are
 drastically reduced for the force which simulates
 the polarisation effects in infinite neutron matter.

\end{abstract}


\maketitle

\section{Introduction}
\label{sec:intro}
 The possibility of nuclear superfludity in neutron stars was
 suggested long  ago \cite{migdal,ginzburg}, before the
 first pulsars were actually observed. The first experimental fact
 pointing to the nuclear  superfluidity in neutron
 stars was the  large relaxation times which follow the sudden
 period changes  ( so-called "glitches'")  of pulsar rotation.
 Later on, the mechanism of the glitch phenomenon  itself was
 directly related to the rotational behaviour of the inner crust
 superfluid\cite{anderson}. Thus, according to the present
 models the  glitches are generated  by a catastrophic unpinning
 of the  superfluid vortex lines  from the nuclei  immersed in the inner
 crust of neutron stars.

 Apart from the glitches phenomena, the superfluid properties of the
 inner crust matter have also important consequences on the cooling
 of neutron stars \cite{lattimer}.
 Thus, the heat diffusion from the interior of the star to the nuclear
 surface  can be strongly suppressed in the crust region due to the
 energy gap in the excitation spectrum of the inner crust superfluid.

 A fully microscopic calculation of pairing properties of
 inner crust matter should be based on a bare nucleon-nucleon force
 and should account for  the polarisation effects induced by the nuclear
 medium.
  However, this is a difficult problem which is not yet completely
 solved even for the pure neutron matter \cite{lombardo}.
 In addition to the uncertainty related to the appropriate
 pairing force, for the inner crust one has also to face the
 problem of calculating the pairing properties of a non-uniform
 nuclear system, in which the density is changing significantly
 from the center of the nuclear clusters to the dilute neutron
 matter in which the clusters are immersed.

  The first calculations of the inner crust matter superfluidity
 were done  by employing semiclassical pairing models based on a local
 density  approximation \cite{broglia}. Later on the calculations
 were performed also in the Hartree-Fock-Bogolibov (HFB)  approach
 \cite{barranco}.
 In these works the mean field of the inner crust matter
 was fixed to a Woods-Saxon form and the pairing field was estimated
 by using a Gogny force \cite{gogny} and a bare force (Argonne).
 The HFB calculations
 showed that the semiclassical pairing models overestimate the
 influence of the nuclear clusters on the
 pairing properies of the inner crust matter.

 In this paper we present  the pairing properties of the inner crust
 matter predicted by fully self-consistent HFB
 calculations in which both the mean field and the pairing field
 are calculated starting from effective two-body forces.
 In this way, the effects induced upon the pairing
 correlations by the strong local variations of the particle density
 and of the nucleon effective mass inside the inner crust matter
 are taken into account consistently. Moreover, the self-consistency
 allows us to study in more details how the properties of the nuclear
 clusters are affected by the external neutron gas in which they
 are embedded.

 In the present HFB calculations the pairing correlations are
 evaluated by using
 a density-dependent contact force. The calculations are done with two
 sets of  parameters, fixed to reproduce  the superfluid properties
 of  neutron matter given either by a Gogny force or by
microscopic
 calculations which take into account  polarisation effects
 \cite{wambach,shen}.

  The article is organized as follows. In Section II we present shortly
 the structure of the inner crust matter used in this study and
 we introduce the HFB approach. The results of the
 calculations are presented in Section III. In the first part of this
 section we analyse the density and the mean field distributions
 and in the second part we discuss the pairing properties.
 The summary and the conclusions are given in Section IV.

\section {Inner crust matter and the HFB approach}
 According to standard models the inner crust matter is formed by a lattice
 of neutron-rich nuclei immersed in a sea of unbound neutrons
 and relativistic electrons \cite{pethick}. In the density range from
 neutron drip density  ($\rho_d \approx 1.4 \times 10^{-3} \rho_0$,
 where $\rho_0$=0.16 fm$^{-3}$ is the saturation density)
 to about half the saturation density,
 the nuclear clusters are most  probably spherical and the
 unbound neutrons are in the  $^1S_0$ superfluid phase.
 At higher densities, before the nuclear clusters are dissolved
 in  the uniform matter of the core, other non-spherical nuclear
 configurations  (e.g.,rods, plates,tubes, bubbles) can be
 formed.

 One of the first microscopic studies of the properties of
 inner crust  matter was done in  Ref.\cite{negele}.
 In this reference the distribution  of the baryonic matter in
 the inner crust was
 evaluated by using a  density  functional suggested by
 the density matrix expansion method and adjusted on the
 neutron equation of state of  Siemens-Pandharipande\cite{sp}.
   In these  calculations the pairing correlations and the spin-orbit
  interaction for the neutrons were neglected. The calculations
  were performed in the Wigner-Seitz approximation, i.e.,
  the inner crust matter is replaced by a set
  of non-interacting cells, each cell containing in its center
  a nuclear cluster surrounded by a gas of unbounded neutrons.
  In addition, in each Wigner-Seitz cell are uniformly distributed a
number
  of relativistic electrons equal to the number of  protons
  contained in the nuclear cluster.

 In this study we analyse the superfluid properties of the
 inner crust matter starting from a set of such  Wigner-Seitz
 cells determined in Ref\cite{negele}. More precisely, for a given
 total baryonic density and proton number we calculate in the
 HFB approach what are the mean fields and the pairing fields
 of the nuclear matter distributed in the Wigner-Seitz cells.

 The HFB equations are solved here in the coordinate space.
 The calculations are performed by using  a Skyrme-type force
 for the particle-hole channel and  a zero range pairing force
 in the particle-particle channel.
 In this case the
 HFB equations are local and for a spherically symmetric system
 they reduce to a set of radial equations \cite{bulgac,dobaczewski}:

\beq
\begin{array}{c}
\left( \begin{array}{cc}
h(r) - \lambda & \Delta (r) \\
\Delta (r) & -h(r) + \lambda
\end{array} \right)
\left( \begin{array}{c} \mathfrak{U}_i (r) \\
 \mathfrak{V}_i (r) \end{array} \right) = E_i
\left( \begin{array}{c} \mathfrak{U}_i (r) \\
 \mathfrak{V}_i (r) \end{array} \right) ~,
\end{array}
\label{1}
\eeq
\\
where $U_i$, $V_i$ are the upper and lower components of the
radial HFB wave functions,
$\lambda$ is the chemical potential while
$h(r)$ and $\Delta(r)$ are the mean field hamiltonian and
pairing field, respectively. They depend on  particle density
$\rho(r)$, abnormal pairing tensor $\kappa(r)$, kinetic energy density
$\tau(r)$ ans spin density $J(r)$ defined by:

\beq
\rho(r) =\frac{1}{4\pi} \sum_{i} (2j_i+1) \mathfrak{V}_i^* (r)
\mathfrak{V}_i (r)
\label{2}
\eeq
\beq
\kappa(r) = \frac{1}{4\pi} \sum_{i} (2j_i+1) \mathfrak{U}_i^* (r)
\mathfrak{V}_i (r)
\label{3}
\eeq
\beq
J(r) = \frac{1}{4\pi} \sum_i (2j_i+1)
[j_i(j_i+1)-l_i(l_i+1)-\frac{3}{4}]
\ V_i^2
\label{4}
\eeq
\beq
\tau(r) = \frac{1}{4\pi} \sum_{i} (2j_i+1)
[(\frac{dV_i}{dr}-\frac{V_i}{r})^2 +\frac{l_i(l_i+1)}{r^2} V_i^2 ]
\eeq
\\
\noindent
The summations are over the whole positive-energy quasiparticle spectrum
of the system.
For the unbound quasiparticle states the summations should
be replaced by integrals over the energy \cite{grasso}.
The general expressions of the mean field and  pairing field
on the densities are given in Ref.\cite{dobaczewski}.

 The HFB equations are applied here to the
inhomogeneous matter of a spherical Wigner-Seitz cell, i.e., a nuclear
cluster immersed in a sea of unbound neutrons. Thus, at relatively large
distance from the nuclear cluster the particle density and
the mean field should have constant non-zero values corresponding
to a uniform neutron gas. In order to get this physical
situation we impose at the cell radius $R_c$ the following boundary
conditions for the HFB solutions \cite{negele}: i) even parity
wave functions vanish at $r=R_c$; ii) first derivatives of
odd-parity wave functions vanish at $r=R_c$. With these mixed boundary
conditions at the
cell border the continuous quasiparticle spectrum is discretized
and all the HFB wave functions corresponding to
bound and unbound states are normalized in the Wigner-Seitz cell.

In the HFB calculations we use for the particle-hole channel
the Skyrme effective interaction SLy4 \cite{sly4}.
This interaction has been adjusted to describe properly
nuclei with a large neutron excess as well as
the properties of  neutron matter.
  Due to these constraints one expects that the
 nuclear matter distribution provided by SLy4 is not far from
 the one estimated in Ref.\cite{negele}. Thus, as seen from
 the calculations presented in Ref.\cite{douchin} the force
 SLy4 predicts for the inner
 crust matter almost the same maximum density and proton fractions
 as the density functional employed in Ref.\cite{negele}.

 The pairing field is calculated here with a  density-dependent contact
 force of the following form
\cite{bertsch}:

\beq
V (\mathbf{r}-\mathbf{r^\prime}) = V_0 [1 -
\eta(\frac{\rho}{\rho_0})^{\alpha}]
\delta(\mathbf{r}-\mathbf{r^\prime})
\equiv V_{eff}(\rho(r)) \delta(\mathbf{r}-\mathbf{r^\prime}).
\eeq

\noindent
 With this force the pairing field is local and is given by:

\beq
\Delta(r) = V_{eff}(\rho(r)) \kappa (r).
\eeq

\noindent
 Since the pairing force has a zero range the HFB calculations
 should be performed with an energy cut-off for the quasiparticle
 spectrum.

 For the parameters of the pairing force we  use
 the following values:  $V_{0}$=-430.0 MeV fm$^3$ , $\eta$=0.7, and
 $\alpha$=0.45. With these values and with a cut-off energy
 equal to 60 MeV we reproduce approximately
 the pairing properties of neutron matter given by the
 Gogny force \cite{bertsch,garrido}.
 In subsection 3.2 we  perform also a HFB calculation with a second
 set of parameters for the pairing force adjusted to reproduce
 the gap values of neutron matter predicted by calculations which
 take into account in-medium polarisation effects.

\section{ HFB properties of inner crust matter }
 As we have mentioned in the previous section, the HFB calculations
 will be performed for a set of representative Wigner-Seitz cells
 determined in Ref.\cite{negele}.

 The maximum  density  for the inner crust considered in
 Ref.\cite{negele} is $\rho_{max}$=0.0789 fm$^{-3}$ and corresponds to
 a cell formed by 32  protons and 950 neutrons. Above this density the
 energy per baryon  becomes close to the  value of the uniform
 neutron system and  other non-spherical configurations might be formed.
 The density  region between the neutron drip density $\rho_d$ and
 $\rho_{max}$ is divided in Ref.\cite{negele} into 11 domains. What is
 remarkable is that the cells correspondings to densities smaller than
 $\rho_{max}$ contain only 50 or 40 protons.

 From all the cells with 50 protons we will consider here only
 two representative cells, i.e.,  one with the density
 $\rho$=0.0204 fm$^{-3}$
 and containing 1750 neutrons, and one with the density
 $\rho$=0.00373 fm$^{-3}$ and having 900 neutrons. These
 cells have the maximum and the minimum densities for
  50 protons. Following  Ref.\cite{negele} we denote the cells
 like a nucleus with Z protons and N neutrons. Thus, the two cells
 with Z=50 will be denoted by  $^{1800}$Sn and $^{950}$Sn.

 We consider
 also two representative cells with Z=40 protons, namely $^{1500}$Zr
 and $^{500}$Zr. These cells corresponds to the densities
 $\rho$=0.0475 fm$^{-3}$ and $\rho$=0.00159 fm$^{-3}$,
 respectively. The cell $^{1500}$Zr is the highest density cell
 with Z=40 while the cell $^{500}$Zr corresponds to the lowest
 positive binding energy per nucleon.

 For the four cells chosen above we present first
 the predictions of the HFB calculations for the  densities
 and the mean fields of the nuclear clusters and the
 surrounding neutron gas. Then, we analyse how the distorsions
 of the density distributions induced by the clusters affect
 the pairing properties of nuclear matter inside the cells.
\subsection{Density and mean field distributions}
 In finite nuclei there is always a maximum number of neutrons
 which can be bound for a given number of protons. This neutron
 stability limit, which defines the neutron dripline,  has been
 extensively studied theoretically in the recent years. Experimentally
 there are drastic limitations for approaching the neutron dripline
 in the laboratory since the neutron-rich nuclei are quickly beta decaying.
 This is not the case for the neutron-rich nuclei immersed in the inner
 crust since here the beta decay is blocked by the presence of the degenerate
 electron gas uniformly distributed throughout the baryonic
 matter. Consequently, inside the inner crust the nuclei can bind
 more neutrons than the nuclei in the vacuum. In addition, their density
 and mean field can change significantly due to the presence of the
 surrounding neutron gas.

 These changes will be analysed first for the case of the Wigner-Seitz
 cells having 50 protons. The HFB calculations with the SLy4 force
 and the pairing interaction mentioned in Section II predict for the
 isolated Sn isotopes the  dripline for two-neutron  separation
 at N=126. Adding two more neutrons to $^{176}$Sn the nuclear
 system in the vacuum is losing about 1 MeV from its binding.

\begin{figure}[h]
\begin{center}
\includegraphics*[scale=0.35,angle=-90.]{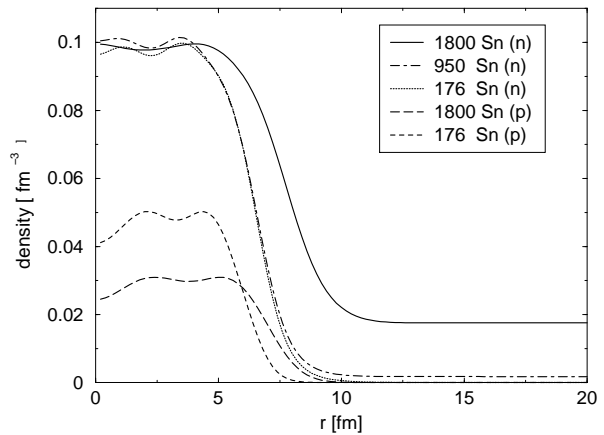}
\caption{Neutron and proton densities for the Wigner-Seitz cells
with 50 protons and for the dripline nucleus $^{176}$Sn. The cells
to which the neutron densities (n) and proton densities (p)
correspond are denoted by the total number of nucleons in the
cell.}
\end{center}
\end{figure}

 In Figs.1-2 the densities and mean fields of the dripline nucleus
 $^{176}$Sn are compared to the corresponding quantities of the nuclear
 cluster immersed in the neutron gas. For the cell $^{950}$Sn one can
  see that the nuclear
 cluster and the dripline nucleus $^{176}$Sn have similar properties.
 Moreover, the number of the bound neutrons inside the cluster is also
 equal to N=126.
 The situation is rather different for the cell $^{1800}$Sn where
 the density of the outer neutron gas is  much higher.
 Thus, it can be seen that the neutron density and the
 neutron mean field profiles have an extended "surface" before
 they reach the constant  values corresponding to the neutron gas.
 In spite of these significant modifications of the density
 and the mean field, the number of bound neutrons inside the cluster,
 i.e., the neutrons with single-particle energies below the constant
 value of the mean field corresponding to the neutron gas, is
 still equal to N=126. On the other hand, if one integrates the neutron
 density up to $r$ = 12 fm, where approximatively the density profile
 is becoming constant, one finds about 300 neutrons. Thus, it appears
 that the large surface is generated in fact by the unbound neutrons
 which are partially localised in the cluster region due to the
 interaction with the bound nucleons.

\begin{figure}[h]
\begin{center}
\includegraphics*[scale=0.35,angle=-90.]{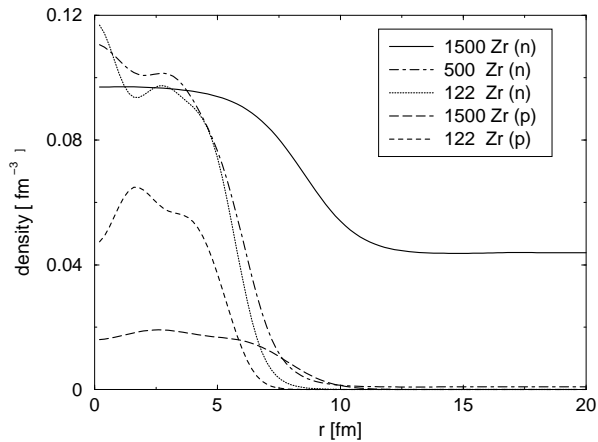}
\caption{The mean fields for the Wigner-Seitz cells with 50
protons and for  the dripline nucleus $^{176}$Sn. The notations
are the same as in Figure 1.}
\end{center}
\end{figure}

Next, we discuss the properties of the nuclear clusters with
40 protons. The neutron dripline for isolated Zr isotopes is
located at N=82. However, in this case the two-neutron separation
energy curve is crossing zero with a very small slope.
Compared to $^{122}$Zr the binding energy of $^{144}$Zr is
only about 200 keV smaller. This is due to the proximity
of the resonant states $3f_{7/2}$ and $3p_{3/2}$ to the
continuum threshold and the fact that these states can
easily accomodate  more neutrons
without losing too much binding energy \cite{sandulescu}.
Thus, one expects that by immersing the dripline nucleus
$^{122}$Zr in the neutron gas one can easily gain the small
extra energy necessary to bind more neutrons in the cluster.
This is actually the case even for the low-density cell $^{500}$Zr.
The density and the mean field profiles for this cell  are shown
in Figs.3-4. From Fig.3 one can see that although the
density of the outer neutrons is about half compared to the cell
$^{900}$Sn, they change more significantly the density profile of the
dripline nucleus $^{122}$Zr than for $^{176}$Sn. Due to these
changes the resonant states $3f_{7/2}$ and $3p_{3/2}$ become
bound, increasing by 12 the number of bound neutrons inside
the nuclear cluster.

\begin{figure}[h]
\begin{center}
\includegraphics*[scale=0.35,angle=-90.]{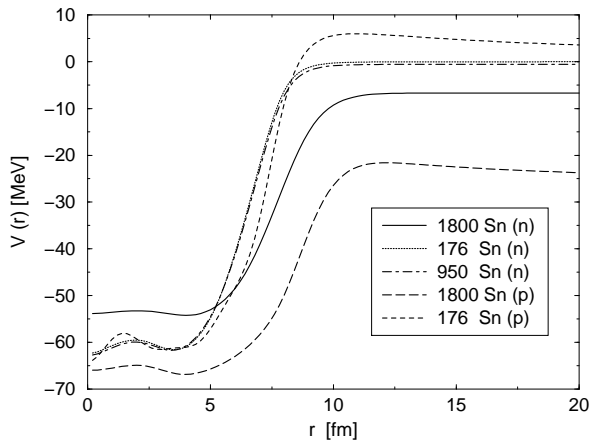}
\caption{Neutron and proton densities for the Wigner-Seitz cells
with 40 protons and for  the dripline nucleus $^{122}$Zr. The
notations are the same as in Figure 1.}
\end{center}
\end{figure}

By increasing the density of the outer neutron gas one expects
further changes of the nuclear cluster formed around 40 protons.
These changes can be seen in Figs.2-3 for the cell $^{1500}$Zr.
However, although the density of the outer neutron gas is now more
than 20 times greater than in the cell $^{500}$Zr, the number of bound
neutrons in the cluster is increasing only by about 10 neutrons.
The total number of neutrons found by integrating the density up to
r=12 fm is about 300 neutrons, as in the cell $^{1800}$Sn.

\begin{figure}[h]
\begin{center}
\includegraphics*[scale=0.35,angle=-90.]{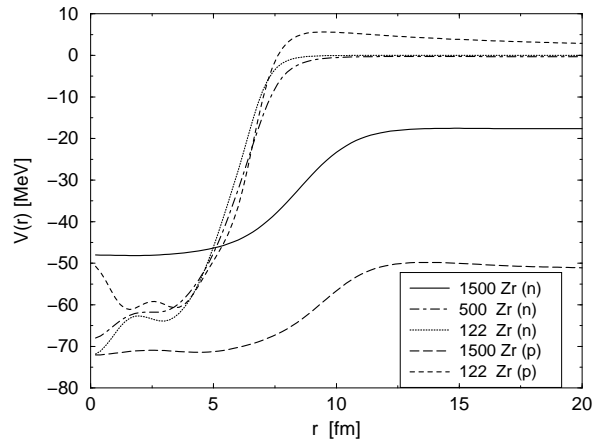}
\caption{The mean fields for the Wigner-Seitz cells with 40
protons and for  the dripline nucleus $^{122}$Zr. The notations
are the same as in Figure 1.}
\end{center}
\end{figure}

In conclusion we find that the nuclear clusters immersed in the
neutron gas keep many features of the dripline nucleus having
the same number of protons. For the cells with large neutron
densities one finds a thick surface-like structure which is
developing in the transition region from the center of the cell
to the neutron gas regime. This surface structure is mainly
formed by  unbound neutrons which are partially kept localised in
the surface region by the interaction with the nucleons bound
inside the nuclear cluster.
\subsection{Superfluid properties}
 As we have seen in the previous subsection, the density of the
neutron matter is changing significantly from the center of the cell,
where the nuclear cluster is located, to its edge filled by the uniform
neutron gas.
As shown by all the microscopic calculations, the pairing gap of the
neutron matter depends strongly on the density. The majority of
the calculations predict that as function of density the pairing gap
has a bell shape form, with the maximum  at a density equal to about
one fifth the saturation density \cite{lombardo}. However, what is
the maximum value of the gap and its detailed density dependence
are still subjects of debate. Thus, the BCS-type calculations
based on bare forces give a maximum gap of about 3 MeV and no
pairing for  nuclear matter at saturation density. Almost
the same maximum gap one gets with a Gogny force, but the gap
falls to a finite value of about 1 MeV at the saturation density.
 On the other hand, all the microscopic calculations which go beyond
the BCS approximation show that the gap is strongly reduced by the
screening and self-energy effects. However, the amount of the
gap supression and even the density dependence of the gap
depend  on the employed approximations. The most recent
calculations point to a maximum gap of about 1 MeV
\cite{wambach,shen,brown}. How this value of the pairing gap
could be reconciled with the pairing gap in finite nuclei is
still an open question.

Since at present it is not yet established what are the
pairing properties of  neutron matter, we perform here
two calculations for the inner crust matter. In one
calculation
we fix the density dependent delta force such as to reproduce
approximately the gap values of  neutron matter provided
by the Gogny force. The parameters of this pairing force, which
we have used above for analysing the density and the mean field
properties of the inner crust matter, are given at the end of
Section II.
In the second calculation, we fix the parameters of the pairing
force so as to obtain for  neutron matter a maximum gap of
about 1 MeV, as in the microscopic calculations which take
into account the screening and the self-energy effects
\cite{wambach,shen}. Since in this case the shape of the
density dependence is unclear, we preferred to change only the
strength of the force, which was reduced to the value
$V_0$=-330 MeV fm$^3$, and to keep the other two parameters
as in the previous calculations. For the cut-off energy
we used in both calculations the same value, i.e.,
$E_c$=60 MeV.

First, we discuss the results given by the pairing interaction
fitted to the Gogny force. The pairing field calculated for the two cells
with 50 protons are shown in Fig.5. For the cell $^{1800}$Sn one can
see that the pairing field is about two times smaller (here and below
we refer to the absolute values of the pairing field) in the center of
the cell compared to the value in the region of the uniform neutron gas.
We can also see that the pairing field is decreasing continuously
towards
the center of the cell.

\begin{figure}[h]
\begin{center}
\includegraphics*[scale=0.35,angle=-90.]{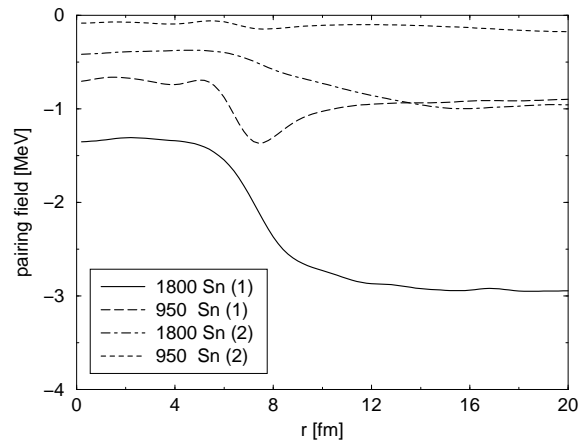}
\caption{The neutron pairing fields for the Wigner-Seitz cells
with 50 protons. The full  and the long-dashed lines correspond to
the pairing fields calculated with the first pairing force while
the  dashed-dotted and the dashed lines correspond to the second
pairing force.}
\end{center}
\end{figure}

 This is not the case for the cell $^{950}$Sn. For this cell
we observe that in passing from the low density region of the
neutron gas towards the high density region of the cluster,
the pairing field is increasing in the intermediate density
region of the cluster surface. This is a manifestation of
the bell shape dependence of the pairing gap on density.
Thus, for the cell $^{950}$Sn the density of the outer
neutron gas is much smaller than the value for which
the gap is maximum, which is reached in the surface
region. In the case of the cell $^{1800}$Sn the
density for which the gap is maximum is already
reached in the neutron gas region.

The behaviour of the pairing field in the cells with 40
protons is rather similar with that found in the two
cells discussed above. As seen in Fig.6, for the
cell $^{500}$Zr we
observe also an increase of the gap in the surface region,
which is now more pronounced than for the cell $^{950}$Sn.
One can also notice that
this significant increase of the pairing field in the
surface region is extending rather deeply towards the
center of the cell. Consequently, the pairing field inside
the nuclear cluster region is not becoming anymore smaller
than in the neutron gas region.

\begin{figure}[h]
\begin{center}
\includegraphics*[scale=0.35,angle=-90.]{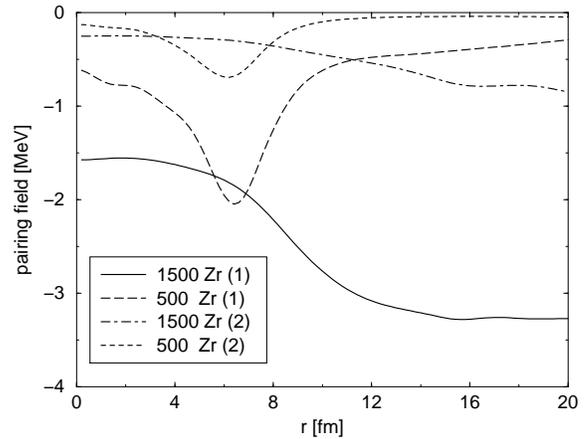}
\caption{Same as Fig.5, for $Z$=40 protons.
}
\end{center}
\end{figure}

 In all the cells we observe that the slope of the pairing field
 is changing very slowly when it is crossing
 the region between the nuclear cluster and the
 uniform neutron gas. This behaviour, sometimes referred to
 as the proximity effect \cite{barranco},
 is related to the large size of Cooper  pairs in
 low-density nuclear matter. The large
 diffusivity of the pairing field in the surface region,
 which is also seen in finite nuclei, is accentuated
 here by the fact that the superfluidity in the surface
 is carried  mainly by the unbound neutrons, which are
 only partially localised at the interface between
 the nuclear cluster and the neutron gas.

Next, we discuss the results corresponding to the
second pairing force which simulates the polarisation
effects. The pairing fields obtained with this
pairing force are shown in Figs.5-6 by short-dashed
and dashed-dotted lines. As expected, the pairing
correlations are reduced strongly for all the cells.
For the high density cells $^{1800}$Sn and $^{1800}$Zr
we can see that, in spite of the drastic supression
of the gap values, the profiles of the pairing fields
are not changing very much compared to the previous
calculations. For the low-density cell $^{950}$Sn we
can see that the pairing field is becoming very small
almost everywhere in the cell. However, in the cell $^{500}$Zr
the pairing field is still showing a significant
increase in the cluster surface and it is almost
vanishing in the neutron gas region.

 Thus, the behaviour of the pairing field in the inner crust
matter is rather complex. As we have seen above, the nuclear
clusters can not only suppress but also enhance  the pairing
field inside the inner crust matter. On the other hand, the
magnitude of the pairing field inside the inner crust depends
very strongly on the scenario used for the pairing properties
of infinite neutron matter.

\section {Summary and Conclusions}

In this paper we have analysed the properties of the inner
crust matter in a self-consistent HFB approach.
In the calculations we used for the particle-hole force
a Skyrme-type interaction, i.e., SLy4, which is able to
describe properly the basic properties of finite nuclei
and infinite neutron matter. As pairing interaction
we have chosen  a density-dependent contact force.
The parameters of the force were fixed so as to
obtain for infinite neutron matter the pairing
properties predicted by two calculations, i.e.,  a BCS
calculation based on Gogny force and a microscopic
calculation based on induced interactions.

The HFB calculations were performed for a set of Wigner-Seitz
cells representative of the structure of the inner crust
matter. First we have studied the density and the mean field
distributions inside the cells and we have analysed the
properties of the nuclear clusters formed in the center
of the cells. Then, we have studied how the pairing field
is modified inside the cells by the presence of the nuclear
clusters. We have thus found that, for the cells corresponding
to high baryonic densities the pairing field is generally
supressed in the region of the nuclear cluster. This is
not the case for the low-density cells where the pairing
field is increasing significantly in the surface region
of the cluster compared to the region of the uniform neutron gas.

The behaviour of the pairing field inside the cells
is rather similar for both pairing forces. However,
the values of the pairing gaps are supressed dramatically
if the pairing force which simulates the polarisation
effects is used in the HFB calculations.

The complex behaviour of the pairing field discussed here
have important consequences on the thermal properties
of the inner crust matter. These aspects will be analysed
in a forthcoming paper \cite{sandulescu2}.

\noindent
{\bf Acknowledgments.}
N. S. acknowledges IPN-Orsay, KTH-Stockholm and the Swedish Programme
for Cooperation in Research and Higher Education (STINT) for financial
support.


\end{document}